\documentclass[aps,preprint]{revtex4}%
\usepackage{amsfonts}
\usepackage{amsmath}
\usepackage{amssymb}
\usepackage{graphicx}%
\providecommand{\U}[1]{\protect\rule{.1in}{.1in}}

\begin{document}
\title[ ]{The Zeeman Effect in Hydrogen Treated in Classical Physics with Classical
Zero-Point Radiation}
\author{Timothy H. Boyer}
\affiliation{Department of Physics, City College of the City University of New York, New
York, New York 10031}

\begin{abstract}
The Zeeman effect for the low resonant energy states of hydrogen is treated
within classical electrodynamics including classical zero-point radiation.
\ The electron is regarded as a classical charged particle in a Coulomb
potential. \ The \textquotedblleft space quantization\textquotedblright\ of
old quantum theory, the Sommerfeld relativistic result, and the Stern-Gerlach
experiment are all considered. \ 

\end{abstract}
\maketitle

\section{Introduction}

Although Faraday looked for the influence of magnetic fields on spectral
lines, the effect was first seen by Zeeman in 1896.\cite{Zeeman} \ The
explanation, involving classical electron theory by Lorentz, earned both
Zeeman and Lorentz the Nobel prize in 1902. \ However, the \textquotedblleft
anomalous\textquotedblright\ Zeeman effect was found by experimenters, and
seemed impossible to explain within classical theory. \ 

Recent work\cite{hydrogen2026} extending classical electrodynamics but
including \textit{classical zero-point energy, relativity, and resonance}, has
considered the hydrogen atom, without imposed magnetic fields. \ In this
present article, we consider the effects of magnetic fields within the same
classical electromagnetic theory. \ We will give classical explanations for
some aspects of the Zeeman effect and of the Stern-Gerlach deflection. \ The
work reported here involves only \textit{classical} electrodynamics and
involves no \textquotedblleft electron spin.\textquotedblright

\section{Effects of a Magnetic Field on Classical Orbits}

\subsection{Model for Hydrogen}

For hydrogen or hydrogen-like systems as described by classical theory, there
is one electron moving in a classical orbit in a Coulomb potential. \ The
orbit will have a vector angular momentum which can be written as a linear
combination involving the unit vectors in the $\widehat{x},~\widehat{y}%
,~\widehat{z}$ directions. \ When classical zero-point radiation is present
with its isotropic but random forces, all orientations of the angular momentum
vector become equally probable. \ However, if a magnetic field along the
$\widehat{z}$-axis is present, then there is a preferred orientation, and the
action variables\cite{GoldsteinAA} $J_{r},~J_{\theta},$ and $J_{\phi}$ for the
electron orbit will assume approximately \textit{resonant} values with the
zero-point radiation relative to the magnetic field direction $\widehat{z}.$ \ 

If a particle orbit is circular, it corresponds to an action value $J_{r}=0$.
\ Because of the \textit{resonance} between the orbital motion and the
zero-point radiation, the circular orbit with total angular
momentum\cite{GoldsteinAA} $J_{2}=J_{\theta}+J_{\phi}$ can assume only certain
orientations relative to the preferred $z$-axis; these orientations
corresponding to the angular momentum along the $\widehat{z}$-axis given by
$J_{\phi}/\hbar=m=-l,-l+1,...,l-1,l$. \ This situation of discrete values for
the classical angle made with the preferred $z$-axis looks like the space
quantization of old quantum theory. \ 

\subsection{Special Case $m=0$}

If the angular momentum lies in one of the coordinate directions
$\widehat{x},~\widehat{y},~\widehat{z}$, then the orbital motion will be (for
an observer looking down the $\widehat{z}$-axis from positive infinity toward
the origin for a \textit{positive} charge) in the counter-clockwise direction
for $\widehat{z},$ the clockwise direction for the $-\widehat{z}$-direction,
or with the angular momentum vector parallel to in the $xy$-plane for the
$\widehat{x}$ or $\widehat{y}$ directions. \ All the other orientations lie at
intermediate orientations for the orbit.

However, there is one special case, occurring for the hydrogen ground state
where $l=1,$ or for any other atom where \textit{one} outer electron has $m=0$
above a spherically-symmetric core. \ The orientation with $m=0$ is excluded,
because it requires $J_{r}=0$ for an orbit of non-zero dimensions while the
orbit is precessing. \ Such a situation is not possible if the orbit is
maintained by \textit{resonance} between orbital motion and classical
zero-point radiation. \ If the circular orbit of non-zero radius has its
angular momentum oriented parallel to the $xy$-plane, then the radius is
surely changing as the charged particle goes around its orbit and the orbit
precesses, so that $J_{r}\neq0.$ \ The orbit must be precessing because the
angular momentum is perpendicular to the magnetic field. \ The Larmor
precession of the orbit provides both a non-zero constant value of the
$z$-component of angular momentum $J_{\phi}$, and also a variation in the
value of the radial motion, so $J_{r}\neq0.$ Thus the case $J_{r}=0$ and $m=0$
is excluded for an orbit of finite radius having \textit{resonance} between
the orbital motion and zero-point radiation. \ \ 

\section{The Zeeman Effect in Hydrogen}

\subsection{Ground State}

The ground state for hydrogen was given in old quantum theory\cite{Bohr1913}
as Bohr's lowest energy level $n=1$ in
\begin{equation}
U=-\frac{M_{0}e^{4}}{2\left(  n\hbar\right)  ^{2}}. \label{Bohreq}%
\end{equation}
And in classical electromagnetism with classical electromagnetic zero-point
radiation, we expect that in a magnetic field with Bohr's $n=1,$ it will have
$l=1,m=\pm1$. \ There are only two orbits resonant with classical zero-point
radiation, one clockwise and one counter-clockwise. \ The orientation
corresponding to $m=0$ is not allowed, as explained in the paragraph above.
\ In the presence of the magnetic field, motion in one direction increases the
velocity of the charged particle thereby increasing the particle's kinetic
energy, and motion in the other direction decreases the kinetic energy
compared to the situation where there is no magnetic field. \ For small
magnetic fields $B,$ the radius of the equilibrium circular orbit remains
unchanged.\cite{Griffiths271} \ The centripetal equation of motion is%
\begin{equation}
M_{0}\frac{\left(  \pm v_{0}+\Delta v\right)  ^{2}}{r_{0}}=-\frac{Ze^{2}%
}{r_{0}^{2}}+e\frac{\left(  \pm v_{0}+\Delta v\right)  }{c}\Delta B.
\end{equation}
\ Accordingly, the change in nonrelativistic kinetic energy of the electron is
given by%
\begin{equation}
\Delta KE\approxeq\frac{1}{2}M_{0}\left[  \pm2v_{0}\left(  \Delta v\right)
\right]  \approxeq\pm\frac{ev_{0}B}{c}.
\end{equation}
In the situation where the charged particle is slowed (because the charge is
moving contrary to the direction of a free electron), it will move near the
fringes of the orbit required by the Coulomb potential. \ This situation seems
like that given in a discussion of diamagnetism.\cite{diamag}

In the absence of the Coulomb potential, there is only one direction for the
\textit{free} electron, that given by Newton's second law for a
nonrelativistic charged particle%

\begin{equation}
\frac{d\mathbf{p}}{dt}=-M_{0}\frac{v^{2}}{r}\widehat{r}=e\frac{\mathbf{v}}%
{c}\times\mathbf{B}%
\end{equation}
giving
\begin{equation}
\frac{v}{c}=\frac{eBr}{M_{0}c^{2}}.
\end{equation}

\section{Zeeman Effect for Excited States of Hydrogen}

\subsection{Fine Structure of Hydrogen}

If a magnetic field is present when an electron changes from one resonant
excited state to some other state, the electron's energy change will involve
the energy of the resonant excited state. \ Such energy levels will depend on
the fine structure of the hydrogen energy levels. \ A \textit{classical}
approach to the fine structure leads to the Zeeman effect in hydrogen and in
alkali metals involving ideas very different ideas from those in quantum
theory. \ 

The fine structure calculated for the hydrogen energy levels follows the
result obtained by Sommerfeld\cite{Sommerfeld} in old quantum theory, but now
reinterpreted in terms of resonant behavior for a charged particle in
classical zero-point radiation. \ We will work with the classical relativistic
expression given in Goldstein's text on classical mechanics\cite{GoldsteinS}%
\begin{equation}
U(J_{3},J_{2,}J_{1})=M_{0}c^{2}\left[  1+\left(  \frac{e^{2}/c}{J_{3}%
-J_{2}+\sqrt{J_{2}^{2}-\left[  e^{2}/c\right]  ^{2}}}\right)  ^{2}\right]
^{-1/2}. \label{UJJJ}%
\end{equation}
Expanding in $e^{2}/\left[  J_{3}c\right]  $ but suppressing the $e^{2}/c$, we
find the approximation%
\begin{align}
&  U(J_{3},J_{2,}J_{1})=M_{0}c^{2}\left[  1+\frac{1}{J_{3}^{2}}\left(
\frac{1}{1-\left(  J_{2}/J_{3}\right)  +\sqrt{\left(  J_{2}/J_{3}\right)
^{2}-1/J_{3}^{2}}}\right)  ^{2}\right]  ^{-1/2}\nonumber\\
&  =M_{0}c^{2}\left[  1-\frac{1}{2J_{3}^{2}}\left\{  \left(  1+\frac{1}%
{J_{3}J_{2}}+...\right)  +\right\}  +\frac{3}{8}\frac{1}{J_{3}^{4}}+..\right]
\end{align}
or, restoring the $e^{2}/c$%
\begin{equation}
U=M_{0}c^{2}-\frac{M_{o}c^{2}}{2}\left(  \frac{e^{2}}{J_{3}c}\right)
^{2}-\frac{M_{o}c^{2}}{2}\left(  \frac{e^{2}}{J_{3}c}\right)  ^{2}\frac{J_{2}%
}{J_{3}}\left(  \frac{e^{2}}{J_{2}c}\right)  ^{2}+\frac{3}{8}M_{0}c^{2}\left(
\frac{e^{2}}{J_{3}c}\right)  ^{2}\left(  \frac{e^{2}}{J_{3}c}\right)  ^{2}
\label{ClassFS}%
\end{equation}
The approximate result in Eq. (\ref{ClassFS}) is that obtained within
classical mechanics for a particle in a Coulomb (or Kepler) potential. \ 

As was shown in the previous article,\cite{hydrogen2026} the excited states of
hydrogen correspond to resonance between the particle orbital motion and
classical electromagnetic zero-point radiation. \ If we write, $J_{3}%
=n_{3}\hbar$ and $J_{2}=n_{2}\hbar,$ then the approximation in Eq.
(\ref{ClassFS}) becomes%
\begin{equation}
U\left(  n_{3},n_{2},n_{1}\right)  =M_{0}c^{2}-\frac{M_{o}c^{2}}{2n_{3}^{2}%
}\left(  \frac{e^{2}}{\hbar c}\right)  ^{2}-\left[  \frac{M_{o}c^{2}}%
{2n_{3}^{2}}\left(  \frac{e^{2}}{\hbar c}\right)  ^{4}\right]  \left[
\frac{1}{n_{3}n_{2}}-\frac{3}{4n_{3}^{2}}\right]  . \label{SommEq}%
\end{equation}
The first term in Eq. (\ref{SommEq}) is the relativistic \textit{rest energy}
of the charged particle, the second term is the \textit{Bohr energy}, and the
third term is the \textit{fine-structure energy} of hydrogen. \ As is often
mentioned in the physics literature, the Sommerfeld result in old quantum
theory agrees with that of the Dirac equation.\cite{Griffiths274}

\subsection{Zeeman Effect for Bohr's $n=2$ Levels}

For Bohr's $n=2$ in Eq. (\ref{Bohreq}), we have $J_{3}=2\hbar.~$However, now
we can have either $J_{r}=\hbar$ and $J_{2}=\hbar,$ or we can have $J_{r}=0$
and $J_{2}=2\hbar$. \ The eccentricity of the \textit{nonrelativistic}
classical orbit is given by%
\begin{equation}
\epsilon=\sqrt{1-\left(  \frac{J_{2}}{J_{3}}\right)  ^{2}}.
\end{equation}
Thus the eccentricities are different for these orbits. \ The lower value of
$J_{2},$ where $J_{r}=1$ corresponds to nonrelativistic elliptical
(relativistic rosette) orbits with smaller angular momentum where the
nonrelativistic eccentricity is $\sqrt{3/4}.$ \ For the second orbits where
$J_{3}=J_{2}$, $J_{r}=0,$ the eccentricity is zero, a circular orbit. \ In the
absence of a magnetic field, the orbit can have any orientation. \ If a
magnetic field is present, the classical orbit might be precessing around the
magnetic field direction. \ 

We see from the $1/\left(  n_{3}n_{2}\right)  $ dependence in Eq.
(\ref{SommEq}) that the levels where the resonant value of $n_{2}=l$ is
smaller have lower (deeper) energy levels compared to those where $n_{2}=l$ is
larger. \ Thus we expect that the the energies for $n_{3}=2,l=1$ are lower
than those for $n_{3}=2,l=2.$ \ There are two lower energy levels,
corresponding to $n_{3}=1,l=1,m=\pm1$,~and these will exhibit the same doublet
sort of splitting in a magnetic field as found for the ground state. \ On the
other hand, the higher levels, corresponding to $n_{3}=2,l=2,m=\pm2,\pm1,$
will exhibit a four-way splitting in a weak magnetic field. \ Remember that
the orientation $m=0$ is excluded by the classical resonant analysis. \ 

As the magnetic field becomes larger, the importance of the Coulomb potential
decreases. \ However, I believe that the Paschen-Back effect does not
correspond to a free electron showing Landau levels. \ The Coulomb potential
is long range. \ 

In our \textit{classical} electromagnetic analysis, the \textit{electron is
simply a point charge}, and there is no such thing as\textit{ }a\textit{
spin-orbit interaction}. \ In contrast, quantum theory regards the fine
structure as arising from two separate effects, the relativistic correction
and the spin-orbit coupling. \ In his quantum text, Griffiths
notes\cite{Griffiths274} \textquotedblleft It is remarkable, considering the
totally different physical mechanisms involved, that the relativistic
correction and the spin-orbit coupling are of the same order $(E_{n}%
^{2}/Mc^{2}).$\textquotedblright\ \ In the classical view, the situation is
not remarkable at all and has nothing to do with \textquotedblleft spin-orbit
coupling.\textquotedblright\ \ It is simply the result of classical
electrodynamics. \ \ We notice from the original relativistic expression, Eq.
(\ref{UJJJ}), that $J_{2}$ cannot vanish since it appears in a square root
with a subtraction.

\section{Stern-Gerlach Experiment}

In the Stern-Gerlach experiment\cite{SG} of 1922, it was found that a beam of
silver atoms was deflected into two separate paths to the collection plate.
\ Later, in 1927, the experiment was repeated with hydrogen atoms by Phipps
and Taylor.\cite{PT} \ Again, at low temperatures, exactly two paths to the
collection plate were found. \ In both cases, the ionized atoms have a charge
valence of $+1$. \ \ Therefore, we believe that there is one outer electron
which can interact with a magnetic field. \ The external magnetic field at the
atom will provide a preferred direction and will split the charged-particle
orbits depending upon what direction (whether clockwise of counter-clockwise)
of motion is involved compared to the magnetic field direction. \ Each charge
particle orbit will have two possible directions of revolution. \ 

\section{Closing Comments}

In this article, we extend\cite{B1975} the work on classical electrodynamics
which includes classical electromagnetic zero-point radiation. \ The analysis
is entirely classical electromagnetic. \ It appears that some more of the
aspects of atomic physics can be explained classically without the need to
invoke quantum notions. \ The Zeeman effect for hydrogen can be discussed in
terms of classical electrodynamics without the need for \textquotedblleft
electron spin.\textquotedblright\ \ \ Indeed for the lower levels of hydrogen,
\textquotedblleft spin\textquotedblright\ seems to refer to the direction of
motion of an electron within its classical orbit. \ 

Also, we must be sure that the orbital motion will allow resonance with
classical zero-point radiation. \ The Sommerfeld relativistic result for
hydrogen can be understood as arising from classical theory where the action
variables of old quantum theory take integer values because of resonance
between the orbital motion and zero-point radiation. \ Classical resonance
seems to explain the \textquotedblleft space quantization\textquotedblright%
\ of old quantum theory. \ Furthermore, the Stern-Gerlach experiment seems to
be explained within classical electromagnetic theory.

March 1, 2026 \ \ \ \ \ \ \ \ \ \ \ Zeeman2.tex


\begin{thebibliography}{99}                                                                                               %


\bibitem {Zeeman}P. Zeeman, \textquotedblleft On the Influence of Magnetism on
the Nature of Light Emitted by a Substance,\textquotedblright\ Phil. Mag.
\textbf{43}, 226-239 (1897).

\bibitem {hydrogen2026}T. H. Boyer, \textquotedblleft Relativistic Hydrogen in
Classical Electrodynamics with Classical Zero-point
Radiation,\textquotedblright\ to be submitted for publication.

\bibitem {GoldsteinAA}H. Goldstein, \textit{Classical Mechanics }2nd edn,
(Addison-Wesley, Reading, MA 1981), p. 575-578. \ We are following Goldstein's notation.

\bibitem {Bohr1913}N. Bohr, \textquotedblleft On the Constitution of Atom and
Molecules, Part 1,\textquotedblright\ Philos. Mag. \textbf{26}, 1-25 (1913).

\bibitem {Griffiths271}D. J. Griffiths, \textit{Introduction to
Electrodynamics} 5th edn (Cambridge U. Press, Cambridge 2024), pp. 273-274.

\bibitem {diamag}T. H. Boyer, \textquotedblleft Diamagnetic behavior in random
classical radiation,\textquotedblright\ Am. J. Phys. \textbf{87},915-923 (2019).

\bibitem {Sommerfeld}A. Sommerfeld, \textquotedblleft Zur Quantentheorie der
Spektrallinien,\textquotedblright\ Annalen der Physik, \textbf{356}, 1--94
(1916). \ 

\bibitem {GoldsteinS}See ref. 3, p. 498.

\bibitem {Griffiths274}D. J. Griffiths, \textit{Introduction to Quantum
Mechanics }2nd ed. (Pearson Prentice Hall, Upper Saddle River, NJ 2005), p.
274-276. \ 

\bibitem {SG}W. Gerlach and O. Stern, \textquotedblleft Der experimentelle
Nachweis der Richtungsquantelung im Magnetfeld\textquotedblright\ Zeitschrift
f\"{u}r Physik \textbf{9}, 349--352 (1922). \ 

\bibitem {PT}T. E. Phipps and J. B. Taylor, \textquotedblleft The Magnetic
Moment of the Hydrogen Atom,\textquotedblright\ Physical Review \textbf{29},
309--320 (1927).

\bibitem {B1975}T. H. Boyer, \textquotedblleft Random electrodynamics: The
theory of classical electrodynamics with classical electromagnetic zero-point
radiation,\textquotedblright\ Phys. Rev. D \textbf{11}, 790-808 (1975).
\end{thebibliography}
\end{document}